\documentclass[conference]{IEEEtran}
\IEEEoverridecommandlockouts

\usepackage{cite}
\usepackage{amsmath,amssymb,amsfonts}
\usepackage{algorithmic}
\usepackage{graphicx}
\usepackage{textcomp}
\usepackage{xcolor}
\usepackage{enumerate}
\usepackage{booktabs}
\usepackage{pifont}
\usepackage{array}
\usepackage{subcaption}
\usepackage[T1]{fontenc}
\usepackage[utf8]{inputenc}
\newcommand{\cmark}{\ding{51}}   

\def\BibTeX{{\rm B\kern-.05em{\sc i\kern-.025em b}\kern-.08em
    T\kern-.1667em\lower.7ex\hbox{E}\kern-.125emX}}
\usepackage[hidelinks]{hyperref}    
    
\begin{document}

\title{DeCEAT: Decoding Carbon Emissions for AI-driven Software Testing}

\author{\IEEEauthorblockN{\textsuperscript{} Pragati Kumari}
\IEEEauthorblockA{\textit{Electrical and Software Engineering} \\
\textit{University of Calgary}\\
Calgary, Canada \\
pragati.kumari@ucalgary.ca}
\and
\IEEEauthorblockN{\textsuperscript{} Novarun Deb}
\IEEEauthorblockA{\textit{Electrical and Software Engineering} \\
\textit{University of Calgary}\\
Calgary, Canada \\
novarun.deb@ucalgary.ca}
}

\maketitle


\begin{abstract}
The increasing use of language models in automated software testing raises concerns about their environmental impact, yet existing sustainability analyses focus almost exclusively on large language models. As a result, the energy and carbon characteristics of small language models (SLMs) during test generation remain largely unexplored. To address this gap, this work introduces the DeCEAT framework, which systematically evaluates the environmental and performance trade-offs of SLMs using the HumanEval benchmark and adaptive prompt variants (based on the Anthropic template). The framework quantifies emission and time-aware behavior under controlled conditions, with \texttt{CodeCarbon} measuring energy consumption and carbon emissions, and unit test coverage assessing the quality of generated tests. Our results show that different SLMs exhibit distinct sustainability strengths: some prioritize lower energy use and faster execution, while others maintain higher stability or accuracy under carbon constraints. These findings demonstrate that sustainability in the generation of SLM-driven tests is multidimensional and strongly shaped by prompt design. This work provides a focused sustainability evaluation framework specifically tailored to automated SLM-based test generation, clarifying how prompt structure and model choice jointly influence environmental and performance outcomes.
\end{abstract}


\section{Introduction}
The landscape of software development is undergoing a rapid transformation, characterized by the pervasive proliferation of Language Models (LMs) across various scales—from multi-billion parameter Large Language Models (LLMs) to highly optimized Small Language Models (SLMs), often deployed through techniques like quantization and knowledge distillation \cite{zadenoori2025model}. This advancement has fueled the extensive integration of Generative AI (GenAI) into Software Engineering (SE) activities, particularly in code generation, debugging, and automated testing, promising substantial gains in developer productivity and efficiency \cite{belcak2025slms}. Despite this clear functional benefit and widespread adoption, there is still very limited research rigorously quantifying the environmental impact—specifically carbon emissions (gCO$_2$e), energy use (kWh) and computational overhead—associated with leveraging these models within the software lifecycle \cite{ashraf2025energyaware}. Our research is therefore motivated by this critical gap, with the objective of providing the essential data necessary to develop sustainable AI-driven software engineering practices and informed model selection.

This paper focuses specifically on the Software Testing phase, particularly functional unit testing ~\cite{zhang2025llmunittesting}. The creation of automated tests involves not only the identification of relevant test cases but also the generation of runnable test scripts for frameworks like Selenium\footnote{https://www.selenium.dev/}, JUnit\footnote{https://junit.org/} or Pytest\footnote{https://pypi.org/project/pytest/}. This is a critical area where Generative AI is seeing increasing adoption, especially for high-volume needs such as Regression Testing ~\cite{ma2024prompt}. However, this efficiency requires a closer examination of its environmental cost. Therefore, the core problem addressed in this research is quantifying the environmental impact of using language models to automate the generation of test scripts. Due to the complexity and variability inherent in large-scale non-functional assessments ~\cite{almonte2025automated}, our exploration is consciously scoped, with integration and system testing (including non-functional properties) being beyond the current scope of this research. 

To address the research problem, we propose the DeCEAT framework (Decoding the Carbon Emission of AI-driven Unit Testing of Software modules). This framework utilizes the \texttt{HumanEval} dataset, a recognized benchmark for measuring language model performance in software testing, focusing on 164 Python code modules. Based on the intuitive premise that they have smaller carbon footprints, we focus exclusively on \text{Small Language Models (SLMs)} with $\leq 8$ billion parameters. We have used---\begin{enumerate}
    \item \texttt{Phi-3.5-mini-instruct} (referred to as \texttt{Phi-3.5-mini})
    \item \texttt{Qwen2.5-1.5B-Instruct} (referred to as \texttt{Qwen2.5-1.5B})
    \item \texttt{deepseek-coder-7b-instruct-v1.5} (referred to as \texttt{deepseek-coder-7b})
    \item \texttt{Mistral-7B-Instruct-v0.3} (referred to as \texttt{Mistral-7B})
    \item \texttt{Meta-Llama-3-8B-Instruct} (referred to as \texttt{Llama-3-8B})
\end{enumerate}
--- for our explorations. To further minimize computational overhead, we employ \textit{quantization} as a model compression technique, specifically investigating whether \textit{4-bit or 8-bit} quantized versions of SLMs are capable of generating adequate test scripts for unit testing. Our approach explores various \textit{prompt engineering techniques}, creating a spectrum of prompts ranging from unstructured (overly simplified) to highly structured, which are then used to instruct the quantized SLMs to generate test scripts for the \texttt{HumanEval} modules. The quality of the generated test scripts is assessed by checking \text{code coverage} achieved using \texttt{coverage.py}. Our initial results show very high accuracy (greater than $90\%$ coverage) in the generated test scripts. Once this high-accuracy capability is established, we proceed to the core sustainability evaluation: measuring the \textit{carbon emission ($\text{gCO}_2\text{e}$), energy consumption ($\text{kWh}$),} and \textit{duration ($\text{s}$)} of these inference tasks using the \texttt{CodeCarbon} plugin. Finally, the DeCEAT framework leverages these raw statistics from \texttt{CodeCarbon} to derive essential \textit{sustainability metrics} that capture the overall environmental impact and importance of using quantized SLMs for automated unit test generation.
The principal contributions of this work to the automated software testing and sustainability communities are multi-fold:
\begin{enumerate}
\item \textbf{\textit{DeCEAT Framework:}} We introduce a novel and fully reproducible evaluation framework designed specifically to assess the environmental impact of prompt-driven test script generation using quantized SLMs.
\item \textbf{\textit{Sustainability Guidance:}} We provide metric-based guidance for practitioners, enabling the selection of optimal and sustainable prompt–model configurations for real-world unit testing deployment scenarios.
\item \textbf{\textit{Novel tradoff Metrics:}} We propose a collection of derived metrics— the \textit{Stability Index }(SI), the \textit{Green Quality Index} ($\text{GQI}$), \textit{Sustainable Coverage Velocity} ($\text{SCV}$), \textit{Sustainable Velocity Index} ($\text{SVI}$), and the \textit{Green $\text{F}_{\beta}$ Score} ($\text{GF}_{\beta}$)\textemdash that capture performance dynamics and highlight the critical interplay between \textit{test quality and environmental trade-offs} within the SLM-based test generation process.
\end{enumerate}

The remainder of the paper is structured as follows: Section \ref{sec:related} documents the related works and gaps in the existing literature. Section \ref{sec:methodology} presents the DeCEAT framework, followed by Section \ref{sec:experimental_setup} that documents the experimental setup. In the next section (Section \ref{sec:results}), we document our experimental results. Section \ref{sec:threats} discusses the
threats to validity of the proposed framework. Finally, Section \ref{sec:Conclusion} concludes the paper.

\section{Related Work}\label{sec:related}

Recent work has explored sustainability in prompting and code generation. Studies such as \cite{wu2023greenprompting,cappendijk2025generating} introduced prompt-based techniques to reduce emissions at inference time. These works evaluated structured prompts across various language models (e.g., GPT-3, Code Llama, DeepSeek-Coder) and demonstrated that prompt phrasing can impact energy usage without compromising output quality. In addition, the influence of prompting patterns on sustainability and model efficiency has been examined in \cite{oprescu2023prompt}, offering structured perspectives on energy-aware prompt engineering.

Efforts to model and monitor carbon emissions throughout the lifecycle of language models are presented in \cite{deng2024llmcarbon,luccioni2023ml}. The corresponding tools and frameworks, including LLMCarbon and ML Bloom, provide lifecycle-aware and real-time tracking of emissions, enhancing transparency and enabling sustainability-informed development decisions.

The environmental impact of automated test script generation has been assessed in \cite{li2023energy,sharma2023sustainable}. These studies analyzed energy usage patterns using models such as CodeT5, GraphCodeBERT, and quantized SLMs, highlighting key considerations in emission-aware testing. Work such as \cite{fraser2011evosuite} introduced EvoSuite, an evolutionary testing tool targeting branch coverage, though without a focus on energy consumption.

Two key surveys \cite{lago2014systematic,mourao2018green} classified green software metrics and sustainability practices, emphasizing the absence of integrated sustainability metrics in early phases of the software lifecycle and advocating for tool-supported green engineering methodologies. Further extensions to sustainability-focused quality frameworks have been proposed in \cite{kapoor2024green}, incorporating maintainability and energy efficiency into software development processes.

Sustainability in large language models has also been examined from operational, deployment, and educational perspectives. Studies such as \cite{iyer2023cloud,verdecchia2021green,vartziotis2024learn} analyzed real-world usage patterns, prompting strategies, and organizational interventions aimed at improving energy and emissions awareness in cloud-hosted AI and code generation contexts. Broader industry guidance is provided through initiatives such as \cite{gsf2025}, which promote best practices and standards for carbon-aware software development. Foundational analyses of energy and carbon costs in NLP tasks were presented in \cite{strubell2019energy,rashkin2021words}, establishing relationships between batch size, model configuration, task type, and resulting carbon intensity.

While prior studies explore sustainability in prompting, LLM inference, and test generation, none specifically examine the environmental characteristics of small language models during automated test script generation. Existing frameworks largely reflect assumptions aligned with large-model behavior, overlooking SLM-specific aspects such as time-aware emissions, prompt sensitivity, and run-to-run stability. This gap underscores the need for an evaluation approach tailored to SLM inference patterns. The DeCEAT framework addresses this need by offering a multidimensional sustainability assessment designed specifically for SLM-based automated test generation.
\section{The DeCEAT Framework}
\label{sec:methodology}

\begin{figure*}[!ht]
    \centering
    \includegraphics[width=0.7\textwidth]{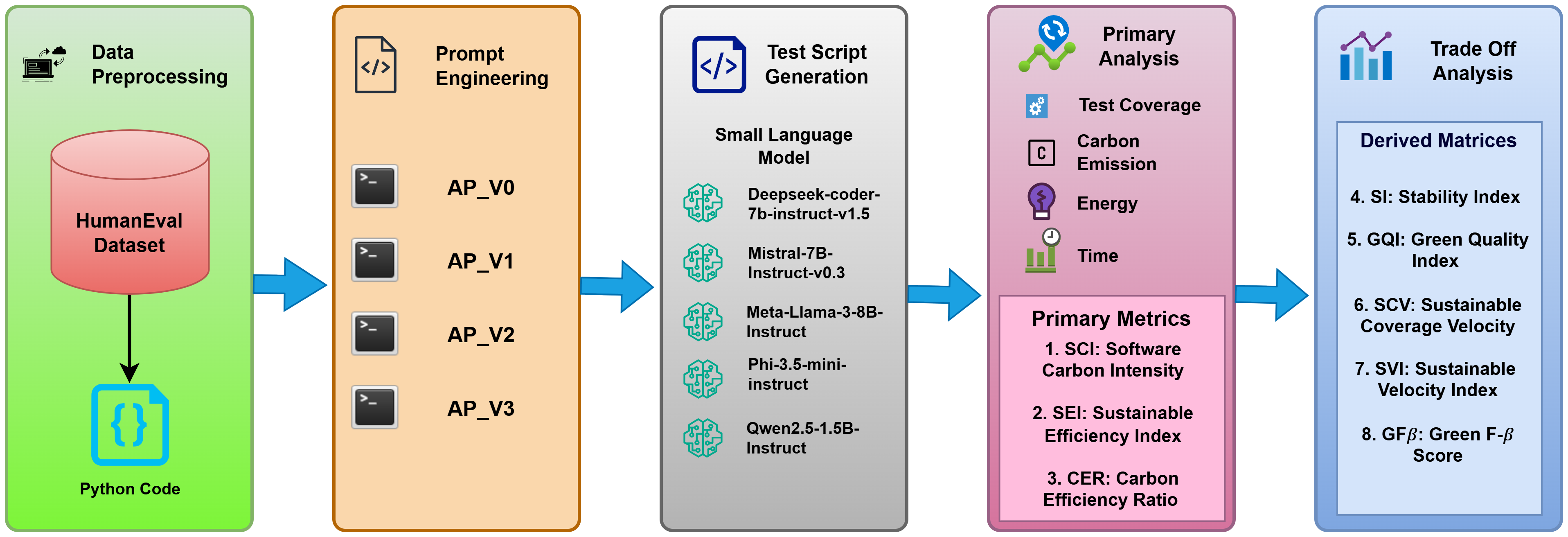}
    \caption{The DeCEAT Framework}
    \label{fig:workflow}
\end{figure*}

To systematically evaluate the sustainability and performance trade-offs of using compact language models for test automation, we propose the DeCEAT framework (Decoding the Carbon Emission of AI-driven Unit Testing). Figure \ref{fig:workflow} illustrates the overall workflow, which is composed of five main phases: (1) \textit{Data Preprocessing}, (2) \textit{Prompt Engineering}, (3) \textit{Test Script Generation}, (4) \textit{Primary Analysis} and (5) \textit{Trade Off Analysis}. The DeCEAT framework is designed to measure the true environmental cost of generating high-quality unit test scripts by employing lightweight, $N$-bit quantized \textit{SLMs} and systematically analyzing their output against established benchmarks. The different phases of the framework are explained in the following sections.

\subsection{Data Preprocessing}


The foundation of our experimental evaluation is the widely adopted HumanEval benchmark \footnote{https://huggingface.co/datasets/openai/openai\_humaneval}, a standardized dataset comprising 164 Python programming tasks used to measure the coding capabilities of Language Models. Each entry in the HumanEval dataset is structured as a record containing five key components:

\begin{enumerate}[(i)]
  \item \textit{task\_id}: a unique identifier for each problem;
  \item \textit{prompt}: the natural language problem statement and the required Python function signature;
  \item \textit{canonical\_solution}: a reference implementation of the correct function logic;
  \item \textit{test}: an assertion-based validation script containing the pre-defined, human-written unit tests; and
  \item \textit{entry\_point}: the callable name of the function to be tested.
\end{enumerate}

For this work, a crucial preprocessing step was necessary to prepare the data for the test script generation and evaluation process. We merged the \textit{prompt} (defining the task and signature) and the \textit{canonical\_solution} (the target code under test) to create a single runnable Python code file for each of the 164 tasks. This unified file represents the \textit{Software Module Under Test} \footnote{https://anonymous.4open.science/r/DeCEAT-1B25}.

This module, combined with its corresponding set of baseline unit tests (the \textit{test} scripts provided by the \texttt{HumanEval} dataset), served as the foundation for the subsequent inference tasks. The Language Models were utilized strictly in a zero-shot inference mode, with no training or fine-tuning performed on this data set. By using this setup, we effectively simulate a realistic automated test synthesis scenario where the Language Models are challenged to infer a comprehensive suite of edge, normal, and invalid test cases from only the minimal contextual information contained within the runnable function code and its original prompt. This step ensures that the generated test scripts are evaluated against the actual function implementation, aligning with practical unit testing workflows.

\subsection{Prompt Engineering} \label{subsec:PrmptVar}

To systematically investigate the influence of instructional complexity on both the quality of the test script and the environmental cost of inference, we designed four progressively constrained prompt templates (AP$_{V_0}$ through AP$_{V_3}$). This structured approach to prompt design was guided by the best practices for instruction tuning and control, particularly drawing from the prompt engineering methodologies proposed by Anthropic \footnote{https://www.youtube.com/watch?v=ysPbXH0LpIE}. Each subsequent variant incrementally introduces additional structure, context, and explicit constraints intended to enhance model behavior and steer the output toward higher-quality unit tests. These constraints serve to mitigate common issues in LLM code generation, such as hallucinations and lack of coverage for edge cases.

\begin{table}[!tbp]
\caption{Comparative analysis of the structural complexity of the prompts derived from the Anthropic template.}
\label{tab:prompt_variants}
\centering
\footnotesize
\setlength{\tabcolsep}{4pt} 
\begin{tabular}{@{} l c c c c @{}}
\toprule
\textbf{Features} & \textbf{AP$_{V_0}$} & \textbf{AP$_{V_1}$} & \textbf{AP$_{V_2}$} & \textbf{AP$_{V_3}$} \\
\midrule
Minimal runnable template            & \cmark & \cmark & \cmark & \cmark \\
Explicit ``expert tester'' persona   &        & \cmark & \cmark & \cmark \\
Structured rules (DOs/DON'Ts)        &        &        & \cmark & \cmark \\
System/user role split               &        &        & \cmark & \cmark \\
Required test methods                &        &        &        & \cmark \\
Tone/format constraints               &        &        &        & \cmark \\
Illustrative example block            &        &        &        & \cmark \\
\bottomrule
\end{tabular}
\end{table}

The comparative structural complexity of the four prompt variants is summarized in detail in Table \ref{tab:prompt_variants}. While the baseline AP$_{V_0}$ serves as a minimal, zero-shot template only containing the runnable code, subsequent versions introduce crucial elements: AP$_{V_1}$ establishes an explicit \textit{"expert tester"} persona; AP$_{V_2}$ layers on explicit structured rules (DOs/DON’Ts) and clearly separates the System/User roles for refined control; and finally, AP$_{V_3}$ (the most constrained variant) introduces explicit requirements for test methods, limits on tone/format, and provides an illustrative example block to ground the model's output format. The exact prompts used for the generation of the test scripts have been provided in our repository \footnote{https://anonymous.4open.science/r/DeCEAT-1B25}.

\subsection{Test Script Generation}

The Test Script Generation phase focuses on the systematic creation of unit tests by the selected language models based on the structured prompts and prepared data modules. For this study, the generation process utilized a specific set of Small Language Models (SLMs), all instantiated under various quantization schemes to maximize efficiency and achieve a low computational footprint during inference. The core evaluation focused on five models divided by their primary quantization level:
\begin{itemize}
    \item[-] \textbf{\textit{8-bit Quantization:}} This scheme was applied to the \texttt{ Phi-3.5-mini} and \texttt{Qwen2.5-1.5B} models.
    \item[-] \textbf{\textit{4-bit Quantization:}} This more aggressive compression was applied to the \texttt{deepseek-coder-7b}, \texttt{Mistral-7B}, and \texttt{Llama-3-8B} models.
\end{itemize}
\noindent In addition to the core five, several models were explored for completeness:
\begin{itemize}
    \item[-] \texttt{TinyLlama/TinyLlama-1.1B-Chat-v1.0} was tested without any quantization, although it yielded weak and unstable output.
    \item[-] \texttt{google/codegemma-7b} was tested using 4-bit quantization; however, while it performed reliably with minimal prompts (AP$_{V_0}$ and AP$_{V_1}$), its performance became limited by system-role support and session timeouts when using highly structured prompts (AP$_{V_2}$ and AP$_{V_3}$).
    \item[-] Similarly, \texttt{google/gemma-2b-it} and \texttt{google/gemma-2-2b-it} faced similar AP$_{V_2}$, AP$_{V_3}$ constraints due to limitations in handling complex role structures and hitting GPU session time limits, making them unsuitable for the full prompt-variant sweep.
\end{itemize}

Each of the prepared code modules is passed to the quantized SLMs using the engineered prompt variants (AP$_{V_0}$ to AP$_{V_3}$). To maintain the scientific rigor of the experiment, all key generation hyperparameters are fixed across all trials to isolate the effects of prompt structure and model quantization. For example, parameters such as the generation \textit{temperature} (set to 0.2), \textit{max\_new\_tokens} (set to 1024), and \textit{batch size} (5 code files per batch) are kept constant. It is important to note that these specific values and constraints are examples and would vary depending on the experimental setup and hardware constraints of other users.

The entire generation process is executed within a controlled environment and is designed to produce a sufficient number of test scripts to allow statistical stability in the subsequent analysis of quality and environmental metrics. The output is a set of generated unit test scripts for each module, prompt, and model configuration, ready for the next phases of quality assessment.

\subsection{Primary Analysis}\label{subsec:PrimAnls}


To assess the environmental impact of Small Language Models ($\text{SLMs}$) in automated test script generation, the proposed framework evaluates a set of \textit{primary metrics} that are derived directly from the fundamental evaluation parameters: \textit{Energy Consumption} ($\text{E}$), \textit{Carbon Emission} ($\text{C}$), and \textit{Test Quality/Code Coverage} ($\text{Q}$). These parameters form the basis for quantifying both the functional utility and the environmental cost of the generated test scripts. This section details these primary metrics and their calculation.

\subsubsection{Software Carbon Intensity (SCI)}

The \textit{Software Carbon Intensity (SCI)} metric, adapted from the Green Software Foundation's standard \cite{gsf2025}, represents the environmental cost of a model's computational activity as the amount of \textit{grams of $\text{CO}_2$ equivalent ($\text{gCO}_2\text{e}$)} emitted per functional run. It is directly proportional to the total energy consumed (by CPU, GPU, and RAM) and can be expressed as:
\begin{equation}
\label{eq:sci}
\text{SCI} \propto E \;\implies\; \text{SCI} = kE,\ \text{where } k = \frac{I}{R}.
\end{equation}

Here, the proportionality constant $k$ is defined by the fixed environmental and operational parameters of the experiment.
\begin{itemize}
    \item $E$ (in $\mathrm{kWh}$) denotes the \textit{total energy consumption} recorded during the entire batch of test generation.
    \item $I$ (in $\mathrm{gCO}_2/\mathrm{kWh}$) is the \textit{carbon intensity} of the power grid where the experiments were carried out.
    \item $R$ is the \textit{number of functional runs} (i.e., the number of individual Python code modules from the \texttt{HumanEval} dataset processed within that batch).
\end{itemize}

\vspace{2pt}
\noindent
\textit{Interpretation:}
$\text{SCI}$ quantifies the carbon cost associated with generating one complete execution cycle (i.e., generating the test script for one code module). A \textit{lower $\text{SCI}$} signifies reduced energy-to-output emissions, providing a direct measure of energy efficiency in the model's operation.

\vspace{2pt}
\noindent
\textit{Relevance to Sustainability:}
$\text{SCI}$ forms the foundation of the sustainability framework, establishing a baseline for emission-aware performance comparison. It quantifies the immediate \textit{carbon footprint of productivity} and aligns with the principles of environmental efficiency under ISO~14067 and the Green Software Foundation guidelines \cite{gsf2025}.

\subsubsection{Sustainable Efficiency Index (SEI)}

The \textit{Sustainable Efficiency Index (SEI)} provides an inverse measure of $\text{SCI}$, indicating the \textit{number of functional executions achievable per gram of $\text{CO}_2$ emitted}.

\begin{equation}
\label{eq:sei}
\text{SEI} = \frac{1}{\text{SCI}}.
\end{equation}

While $\text{SCI}$ captures the environmental cost per output, $\text{SEI}$ emphasizes the return—how effectively computational effort converts into productive inference.

\vspace{2pt}
\noindent
\textit{Interpretation:}
A \textit{higher $\text{SEI}$} indicates greater computational productivity under fixed carbon constraints, linking efficiency directly with the emission economy. It complements $\text{SCI}$ by reflecting the sustainability gain per unit of emission rather than just its cost.

\vspace{2pt}
\noindent
\textit{Relevance to Sustainability:}
$\text{SEI}$ captures the principle of \textit{doing more with less} by translating energy optimization into meaningful output efficiency. It bridges traditional performance evaluation with ecological responsibility, quantifying the \textit{positive sustainability return} from energy-efficient model behavior.




\subsubsection{Carbon Efficiency Ratio (CER)}

The \textit{Carbon Efficiency Ratio (CER)} integrates model output quality with environmental cost, expressing the \textit{percentage of code coverage achieved per gram of emitted $\text{CO}_2$}.

\begin{equation}
\label{eq:cer}
\text{CER} = \frac{Q}{C}
\end{equation}

In this formula:
\begin{itemize}
    \item $Q$ denotes the code coverage achieved by the generated test scripts (expressed as a percentage, \%).
    \item $C$ represents the total carbon emission per prompt execution or functional run (in $\text{gCO}_2\text{e}$).
\end{itemize}

\vspace{2pt}
\noindent
\textit{Interpretation:}
A \textit{higher $\text{CER}$} represents greater output quality per unit of emission, effectively balancing test accuracy with ecological expenditure. This metric extends sustainability assessment beyond simple energy conservation to include reliable performance outcomes.

\vspace{2pt}
\noindent
\textit{Relevance to Sustainability:}
$\text{CER}$ emphasizes \textit{quality-oriented sustainability}, ensuring that emission-efficient models also deliver a consistent and complete test coverage. By correlating coverage ($Q$) with emission data ($C$), it reinforces the idea that carbon efficiency must reflect both energy responsibility and result integrity within software testing workflows.




\subsection{Trade Off Analysis}

This section introduces a collection of derived metrics designed to facilitate a nuanced trade-off analysis between model performance (test quality) and environmental sustainability (carbon and energy expenditure). While the primary metrics established the foundational costs and efficiencies, these derived metrics explicitly capture the interplay and complexity of optimizing both factors simultaneously. By combining core parameters, they provide a powerful quantitative basis for users and researchers to select optimal, high-quality, and environmentally responsible models and prompt configurations for automated test script generation.

\subsubsection{Stability Index (SI)}

The \textit{Stability Index (SI)} measures the \textit{run-to-run consistency} in a model's sustainability performance.

\begin{equation}
\label{eq:si}
\text{SI} = 1 - \frac{\sigma}{\mu}
\end{equation}

In this formula:
\begin{itemize}
    \item $\mu$ denotes the \textit{mean value} of any previously defined primary metric ($\text{SCI}$, $\text{SEI}$ or $\text{CER}$) in multiple inference runs.
    \item $\sigma$ represents the \textit{standard deviation} of the same metric in multiple runs.
\end{itemize}

\vspace{2pt}
\noindent
\textit{Interpretation:}
$\text{SI}$ captures the steadiness of model behavior under repeated execution. A \textit{high $\text{SI}$} indicates a model that maintains uniform emission and energy profiles, while a lower or negative $\text{SI}$ reflects irregular or unstable sustainability patterns.

\vspace{2pt}
\noindent
\textit{Relevance to Sustainability:}
Stability ensures that sustainability gains are not random or context-dependent. Models exhibiting high $\text{SI}$ values deliver predictable and reliable performance\textemdash key to scalable and reliable sustainable $\text{AI}$ deployment. It links directly to environmental reproducibility, confirming that energy efficiency persists in dynamic testing scenarios.

\subsubsection{Green Quality Index (GQI)}
\begin{equation}
\label{eq:GQI}
\text{GQI} = 2 \times \frac{Q \times E}{Q + E}
\end{equation}

\noindent
The \textit{Green Quality Index (GQI)} combines normalized quality ($Q$) and efficiency ($E$) through a harmonic mean, providing a unified view of sustainability progress.

\vspace{2pt}
\noindent
\textit{Interpretation:}
GQI measures the holistic improvement in environmental efficiency and functional quality.  
A higher GQI indicates that a model achieves better energy–quality synergy, effectively balancing emission reduction with reliable output generation.

\vspace{2pt}
\noindent
\textit{Relevance to Sustainability:}  
GQI emphasizes \textit{co-optimization}, where improvements in sustainability are achieved without compromising output performance.  
By integrating emission reduction, energy efficiency, and quality metrics, it represents the comprehensive sustainability maturity of a model—crucial for long-term green AI development.




\subsubsection{Sustainable Coverage Velocity ($\text{SCV}$)}

The \textit{Sustainable Coverage Velocity (SCV)} includes two variants: SCV$_C$, which uses Carbon Emissions (C), and its energy variant SCV$_E$, which uses Energy Consumption (E), both of which integrate time into the sustainability assessment by measuring the \textit{coverage achieved per unit of time and resource consumption}.

\begin{equation}
\label{eq:scv}
\text{SCV}_C = \frac{Q}{T \times C}, \qquad \text{SCV}_E = \frac{Q}{T \times E}
\end{equation}

In these formulas:
\begin{itemize}
    \item $Q, C, E$ have been defined in section \ref{subsec:PrimAnls}, and
    \item $T$ is the execution time (duration of the inference task) in seconds ($\text{s}$).
\end{itemize}

\vspace{2pt}
\noindent
\textit{Interpretation:}
A higher value of $\text{SCV}_C$ or $\text{SCV}_E$ implies a faster coverage progression with a lower time and resource cost, revealing which models deliver faster and cleaner execution cycles.

\vspace{2pt}
\noindent
\textit{Relevance to Sustainability:}
$\text{SCV}$ transforms sustainability into a dynamic perspective, emphasizing how efficiently models utilize time and resources to generate validated outputs. It reflects the practical notion of "\textit{green velocity}"\textemdash achieving maximum computational productivity within minimal carbon and energy budgets.




\subsubsection{Sustainable Velocity Index (SVI)}

The \textit{Sustainable Velocity Index (SVI)} combines four critical sustainability aspects\textemdash coverage, software carbon intensity, execution time, and stability\textemdash into a single normalized metric.

\begin{equation}
\label{eq:svi}
\text{SVI} = \frac{Q}{100} \cdot \frac{1}{1+\hat{\text{SCI}}} \cdot \frac{1}{1+\hat{T}} \cdot (1-\hat{\sigma}), 
\qquad
\hat{x} = \frac{x - x_{\min}}{x_{\max} - x_{\min}}
\end{equation}

In this formula:
\begin{itemize}
    \item $Q$ is the code coverage achieved (\%).
    \item $\hat{\text{SCI}}$ is the normalized Software Carbon Intensity.
    \item $\hat{T}$ is the normalized execution time.
    \item $\hat{\sigma}$ is the normalized average standard deviation of \textit{SCI, SEI,} and \textit{CER} (used as a proxy for instability).
    \item $\hat{x}$ is the normalization function that ensures fair and consistent cross-model comparisons by scaling all metrics to a common range.
\end{itemize}

\vspace{2pt}
\noindent
\textit{Interpretation:}
$\text{SVI}$ quantifies the comprehensive sustainability performance under temporal and variability constraints. A \textit{high $\text{SVI}$} indicates a model that achieves broad coverage ($Q$) with minimal time ($\hat{T}$), consistent stability ($\hat{\sigma}$), and low emissions ($\hat{\text{SCI}}$).

\vspace{2pt}
\noindent
\textit{Relevance to Sustainability:}
The multiplicative formulation of $\text{SVI}$ ensures that deficiencies in one dimension (e.g., long runtime) cannot fully offset strengths in another, promoting \textit{balanced eco-performance}. It rewards models that maintain consistent speed, stability, and emission efficiency simultaneously, reflecting a holistic view of sustainable $\text{AI}$ deployment.

\subsubsection{Green $\text{F}_{\beta}$ Score ($\text{GF}_{\beta}$)}

The \textit{Green $\text{F}_{\beta}$ Score ($\text{GF}_{\beta}$)} synthesizes eco-efficiency ($\text{ECO}$) and code coverage accuracy ($Q$), balancing environmental impact with performance quality.

\begin{equation}
\label{eq:gfbeta}
\text{GF}_{\beta} = \frac{(1+\beta^{2})Q \cdot \text{ECO}}{\beta^{2}Q + \text{ECO}}, 
\qquad 
\text{ECO} = \frac{1}{(1+\hat{\text{SCI}})(1+\hat{T})}
\end{equation}

In this formula:
\begin{itemize}
    \item $Q$ is the code coverage achieved (\%).
    \item $\text{ECO}$ represents the eco-efficiency index, combining normalized $\text{SCI}$ ($\hat{\text{SCI}}$) and normalized execution time ($\hat{T}$), where $T$ is the duration in seconds.
    \item $\beta$ is the \textit{weighting parameter} that modulates the emphasis placed on quality versus eco-efficiency.
\end{itemize}
The resulting mean values of $\text{GF}_{\beta}$ were grouped into two comparative regimes: $\beta < 1$ representing the \textit{eco-efficiency domain} and $\beta > 1$ representing the \textit{quality-oriented domain}.

\vspace{2pt}
\noindent
\textit{Interpretation:}
$\text{GF}_{\beta}$ provides a tunable trade-off between environmental and performance priorities. The values of $\beta < 1$ prioritize eco-efficiency, while $\beta > 1$ prioritize output quality. Higher values in the $\beta < 1$ regime indicate stronger eco-efficiency under time and emission constraints, whereas higher values in the $\beta > 1$ regime signify improved quality retention despite ecological limitations.

\vspace{2pt}
\noindent
\textit{Relevance to Sustainability:}
$\text{GF}_{\beta}$ integrates the dynamics of time, emission and quality into a single composite index, allowing a holistic evaluation of how models optimize resource usage while maintaining accuracy of coverage. It is particularly suited for sustainable systems in the real-world $\text{AI}$, where both energy optimization and quality assurance are essential for the implementation of environmentally responsible models.

\begin{table}[!t]
\caption{Parameter inclusion across sustainability metrics.}
\label{tab:metric_params_flipped}
\centering
\footnotesize
\setlength{\tabcolsep}{6pt}
\renewcommand{\arraystretch}{1.2}
\begin{tabular}{@{} l c c c c l @{}}
\toprule
\textbf{Metric} & \textbf{C} & \textbf{E} & \textbf{Q} & \textbf{T} & \textbf{Interpretation} \\ 
\midrule
SCI           &            & \cmark &          &          & Carbon emission per run \\
SEI           &            & \cmark &          &          & Efficiency per CO$_2$ unit \\
CER           & \cmark     &        & \cmark   &          & Coverage per emission \\
SI            & \cmark     & \cmark & \cmark   &          & Run-to-run stability \\
GQI           &            & \cmark & \cmark   &          & Energy--quality balance \\
SCV$_C$ / SCV$_E$ & \cmark     & \cmark & \cmark   & \cmark   & Time-integrated velocity \\
SVI           & \cmark     & \cmark & \cmark   & \cmark   & Time--carbon--stability index \\
GF$_\beta$    &            & \cmark & \cmark   & \cmark   & Unified eco--quality score \\
\bottomrule
\end{tabular}
\end{table}

\section{Experimental Setup}
\label{sec:experimental_setup}

Building upon the DeCEAT framework described in Section~\ref{sec:methodology}, this section outlines the complete experimental configuration designed to operationalize the framework and evaluate sustainability–performance trade-offs across multiple Small Language Models (SLMs).  
The setup integrates structured prompt variants (AP\textsubscript{V0}–AP\textsubscript{V3}), quantized model executions, and \texttt{CodeCarbon}-based emission tracking, ensuring consistency, repeatability, and comparability across all models and runs.

\subsection{Environment and Tooling}
All experiments were conducted on \textit{Google Colab} using an \textit{NVIDIA T4 GPU} (16\,GB RAM, free tier).  
The experimental environment employed \texttt{transformers}, \texttt{accelerate}, \texttt{bitsandbytes}, \texttt{codecarbon}, and \texttt{tqdm} for inference, quantization, and emission tracking.  
\textit{CodeCarbon} monitored energy consumption (Wh), emissions (gCO\textsubscript{2}eq), and runtime statistics at the batch level.  
A fixed grid carbon intensity factor \textit{I} (gCO\textsubscript{2}/kWh) was used for metric computation, representing the 12-month average value for Alberta obtained from  \textit{Electricity Maps} \footnote{https://app.electricitymaps.com/map/live/fifteen\_minutes}.  
In addition to emissions, CodeCarbon automatically logged execution time, CPU and GPU power draw, memory usage, and process-level energy distribution, all used for subsequent analyses.

\subsection{Run Configuration}
Each model was executed under identical generation parameters to ensure fairness across all adaptive prompt variants (AP\textsubscript{V0}–AP\textsubscript{V3}).  
Each batch contained five functional executions (\textit{R} = 5), after which a single \texttt{CodeCarbon} log was generated to record energy, emission, and runtime values.  
For each model, one CSV file was produced per prompt version, resulting in a total of $20$ \texttt{CodeCarbon} CSV logs ($5$ models × $4$ prompt variants).  
Each CSV contained $33$ runs per prompt, providing sufficient granularity for averaging and subsequent metric computation.
\begin{table}[!h]
\caption{Generation and run parameters.}
\label{tab:params}
\centering
\footnotesize
\setlength{\tabcolsep}{6pt}
\renewcommand{\arraystretch}{1.2}
\begin{tabular}{@{} l l @{}}
\toprule
\textbf{Parameter} & \textbf{Value} \\ 
\midrule
Temperature          & 0.2 \\
max\_new\_tokens     & 1024 \\
Batch size           & 5 code files per batch \\
Model quantization   & 4-bit or 8-bit (as defined in framework) \\
Tracking tool        & CodeCarbon \\
\bottomrule
\end{tabular}
\end{table}

\subsection{Pipeline Execution and Data Aggregation}
Following the DeCEAT workflow (Fig.~\ref{fig:workflow}), the experimental pipeline consisted of five sequential stages:

\begin{enumerate}[(i)]
    \item \textit{Batch execution:} Each model generated test scripts under four prompt variants (AP$_{V_0}$--AP$_{V_3}$), producing one \texttt{CodeCarbon} log per variant.

    \item \textit{Data consolidation:} The 20 CSV logs were merged into a unified dataset containing emission (kgCO$_2$), energy (kWh), execution time (s), and GPU utilization metrics.

    \item \textit{Coverage measurement:} Unit test coverage (\%) was computed using \texttt{coverage.py} and averaged across all four prompt variants for each model.

    \item \textit{Metric input extraction:} The dataset was used to extract key parameters—coverage, emission (converted to gCO$_2$), energy (kWh), and time (s)—that served as inputs for all primary and derived metrics.

    \item \textit{Evaluation phases:} The computed inputs were used to derive the primary analysis metrics (SCI, SEI, CER, SI) and the trade-off metrics (GQI, SCV, SVI, GF$_\beta$).
\end{enumerate}

This structured process ensured uniform experimental conditions, consistent emission factors, and reproducible sustainability comparisons across models.

\subsection{Metric Computation Setup}
\label{sec:metric_setup}
All metric computations were based on the processed data obtained from the $20$ \texttt{CodeCarbon} CSV logs.  
For each model and prompt, emissions were converted from kgCO\textsubscript{2} to gCO\textsubscript{2}, while energy (kWh) and execution time (s) were directly extracted from the logs.  
Average coverage (\%) was computed using \texttt{coverage.py} and consolidated across all prompt variants per model.  

\noindent
Two constants were maintained throughout:
\begin{itemize}
    \item \textit{I}: Grid carbon intensity (gCO\textsubscript{2}/kWh), representing the 12-month Alberta average from \textit{Electricity Maps}.
    \item \textit{R} = 5: Number of code files per batch for averaging across runs.
\end{itemize}

Primary metrics (\textit{SCI}, \textit{SEI}, \textit{CER}) were derived directly, where \textit{SEI} is the inverse of \textit{SCI} and \textit{CER} denotes the ratio of coverage to emission.  
Derived metrics (\textit{SI}, \textit{SCV}, \textit{SVI}, \textit{GF}\textsubscript{$\beta$}) extended these formulations by integrating time, stability, and normalized indices scaled between 0 and 1 for inter-model comparison.  
For \textit{GF}\textsubscript{$\beta$}, six $\beta$ values—0.3, 0.6, 0.9, 1.2, 1.5, and 1.8—were used to evaluate model performance under both eco-efficiency ($\beta < 1$) and quality-oriented ($\beta > 1$) settings.  

The detailed formulae, normalization steps, and complete calculation sheet are available in the public GitHub repository\footnote{https://anonymous.4open.science/r/DeCEAT-1B25}, which provides a step-by-step guide to every metric derivation for full reproducibility.

\section{Results and Discussion}
\label{sec:results}
This section presents the empirical results of the DeCEAT framework and discusses key sustainability trends and trade-offs observed across the evaluated SLMs and prompt variants.

\begin{figure*}[!t]
\centering
\begin{subfigure}{0.32\textwidth}
    \centering
    \includegraphics[width=\linewidth]{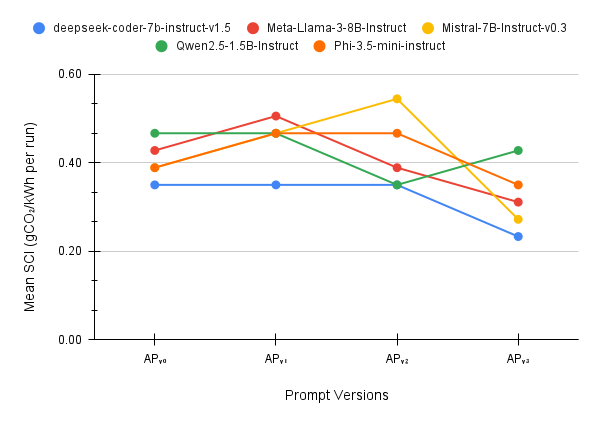}
    \caption{SCI: lower indicates reduced CO$_2$/run.}
    \label{fig:SCI}
\end{subfigure}
\hfill
\begin{subfigure}{0.32\textwidth}
    \centering
    \includegraphics[width=\linewidth]{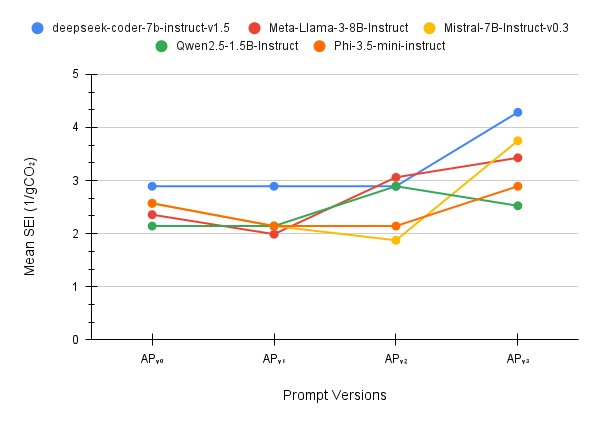}
    \caption{SEI: inverse of SCI; higher is better.}
    \label{fig:SEI}
\end{subfigure}
\hfill
\begin{subfigure}{0.32\textwidth}
    \centering
    \includegraphics[width=\linewidth]{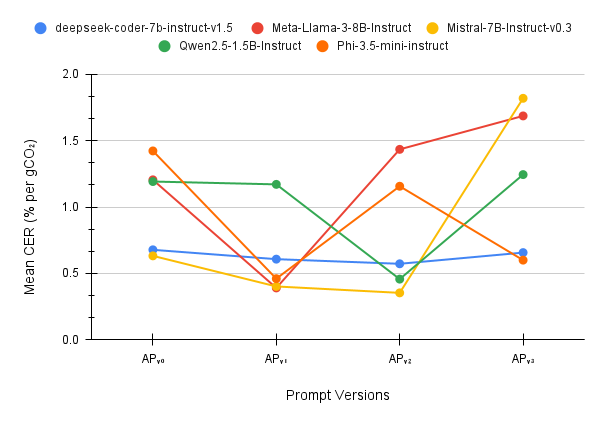}
    \caption{CER: coverage per CO$_2$ unit.}
    \label{fig:CER}
\end{subfigure}

\caption{Primary sustainability metrics across SLMs and prompt variants.}
\label{fig:sci_sei_cer}
\end{figure*}

\subsection{Primary Analysis}
\label{sec:static_results}
To evaluate model-level emission efficiency, three primary sustainability metrics—\textit{SCI}, \textit{SEI}, and \textit{CER}—were analyzed across all adaptive prompt variants AP$_{V_0}$–AP$_{V_3}$.

\subsubsection{Software Carbon Intensity (SCI)}
As defined in \eqref{eq:sci}, \textit{SCI} quantifies the carbon emissions per functional execution. The results are illustrated in Fig.~\ref{fig:SCI}.  

\textit{Observations:}  
\texttt{deepseek-coder-7b} recorded the lowest \textit{SCI} (0.23 at AP$_{V_3}$), confirming its superior energy efficiency and minimal emission footprint.  
\texttt{Llama-3-8B} and \texttt{Mistral-7B} also showed a noticeable decline in \textit{SCI} across prompt versions, with respective drops from 0.43 to 0.31 and 0.39 to 0.27, reflecting improved emission utilization through adaptive prompt refinement.  
\texttt{Qwen2.5-1.5B} maintained relatively higher emission levels (around 0.47), while \texttt{Phi-3.5-mini} displayed stable yet moderate efficiency.  
The overall downward trajectory in \textit{SCI} across AP$_{V_0}$–AP$_{V_3}$ reinforces that structured prompting effectively minimizes redundant computation, leading to lower carbon intensity per execution.

\subsubsection{Sustainable Efficiency Index (SEI)}
As defined in \eqref{eq:sei}, \textit{SEI} represents the number of functional executions achievable per gram of CO$_2$ emitted, serving as the inverse of \textit{SCI}. The results are illustrated in Fig.~\ref{fig:SEI}.  

\textit{Observations:}   
\texttt{deepseek-coder-7b} achieved the highest \textit{SEI} (4.28 at AP$_{V_3}$), confirming its consistent emission-to-output superiority.  
\texttt{Mistral-7B} demonstrated a significant rise in \textit{SEI} from 2.57 at AP$_{V_0}$ to 3.75 at AP$_{V_3}$, establishing its strong efficiency growth across structured prompts.  
\texttt{Llama-3-8B} and \texttt{Phi-3.5-mini} exhibited moderate improvement, reaching 3.43 and 2.89 respectively, while \texttt{Qwen2.5-1.5B} maintained a lower efficiency profile (ranging near 2.5).  
The consistent increase in \textit{SEI} across all models indicates that adaptive prompting promotes sustainable throughput by enhancing computational productivity under fixed emission constrai
nts.

\subsubsection{Carbon Efficiency Ratio (CER)}
As defined in \eqref{eq:cer}, \textit{CER} relates achieved test coverage to total CO$_2$ emissions, measuring the coverage-to-carbon performance trade-off. The results are presented in Fig.~\ref{fig:CER}.  

\textit{Observations:}  
\texttt{Mistral-7B} achieved the highest \textit{CER} (1.82 at AP$_{V_3}$), representing the most efficient coverage generation per emission unit.  
\texttt{Llama-3-8B} followed with a strong \textit{CER} (1.68), while \texttt{Qwen2.5-1.5B} maintained moderate efficiency (1.24).  
\texttt{Phi-3.5-mini} displayed fluctuating coverage-to-carbon performance across prompts, averaging around 1.16, and \texttt{deepseek-coder-7b}, despite being emission-efficient, yielded a relatively lower \textit{CER} (0.65) due to limited coverage variation.  
The clear upward trend in \textit{CER} for \texttt{Mistral-7B} and \texttt{Llama-3-8B} demonstrates that prompt optimization enhances both coverage diversity and sustainability, making them the most balanced performers under adaptive conditions.

\noindent
\textit{Overall Interpretation:}  
Collectively, \textit{SCI}, \textit{SEI}, and \textit{CER} reveal that structured prompt optimization significantly improves emission-aware productivity.  
\texttt{deepseek-coder-7b} remains the most energy-efficient model with the lowest carbon intensity, \texttt{Mistral-7B} leads in combined coverage-to-carbon performance, and \texttt{Llama-3-8B} maintains an ideal balance between emission reduction and coverage consistency.  
These results confirm that adaptive prompting and quantized inference jointly foster greener, more efficient AI-driven test generation workflows.

\subsection{Trade Off Analysis}
\label{sec:dynamic_results}
To assess the reproducibility and overall sustainability gain of each model, two metrcis—\textit{Stability Index (SI)} and \textit{Green Quality Index (GQI)}—were computed as defined in Section~\ref{sec:methodology}.  
These indicators capture both consistency and progressive improvement in emission-aware behavior (Figures~\ref{fig:stability}–\ref{fig:GQI}).

\subsubsection{Stability Index (SI)}
As defined in \eqref{eq:si}, \textit{SI} reflects the overall steadiness of model performance across emission and efficiency metrics. The results are illustrated in Fig.~\ref{fig:stability}.

\textit{Observations:}  
\texttt{Qwen2.5-1.5B} achieved the highest stability (SI = 0.915), indicating exceptional reproducibility across all sustainability metrics and minimal deviation between runs.  
\texttt{deepseek-coder-7b} followed closely (SI = 0.873), maintaining a well-balanced performance across emission, efficiency, and coverage-based parameters.  
\texttt{Phi-3.5-mini} and \texttt{Llama-3-8B} demonstrated moderate stability (SI = 0.758 and 0.727, respectively), suggesting controlled yet noticeable variability under changing prompt versions.  
In contrast, \texttt{Mistral-7B} recorded the lowest steadiness (SI = 0.617), showing higher sensitivity to prompt variation and slightly inconsistent sustainability patterns.  

\vspace{2pt}
\noindent
Overall, models with $\text{SI} \geq 0.6$ exhibit consistent and reproducible sustainability behavior, confirming stable emission–efficiency alignment across prompt variants.  
Among these, \texttt{Qwen2.5-1.5B} and \texttt{deepseek-coder-7b} emerge as the most reliable, showing minimal run-to-run deviation and strong environmental consistency — essential traits for reproducible and trustworthy Green AI benchmarking.


\subsubsection{Green Quality Index (GQI)}
\begin{figure*}[!t]
\centering

\begin{subfigure}{0.32\textwidth}
    \centering
    \includegraphics[width=\linewidth]{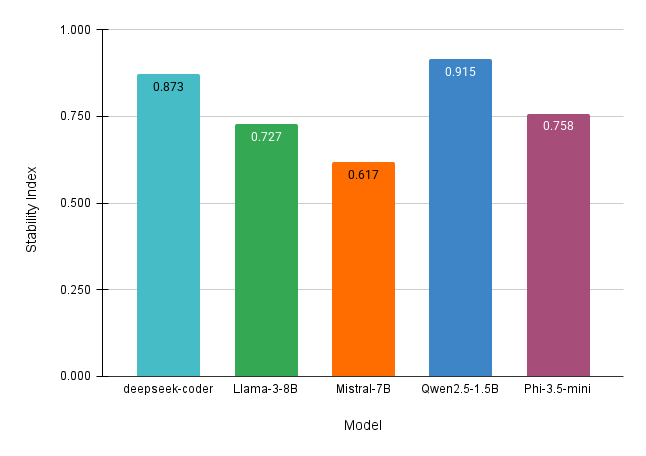}
    \caption{SI}
    \label{fig:stability}
\end{subfigure}
\hfill
\begin{subfigure}{0.32\textwidth}
    \centering
    \includegraphics[width=\linewidth]{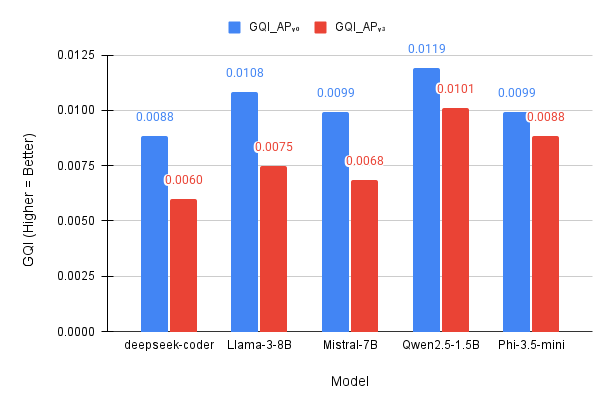}
    \caption{GQI}
    \label{fig:GQI}
\end{subfigure}
\hfill
\begin{subfigure}{0.32\textwidth}
    \centering
    \includegraphics[width=\linewidth]{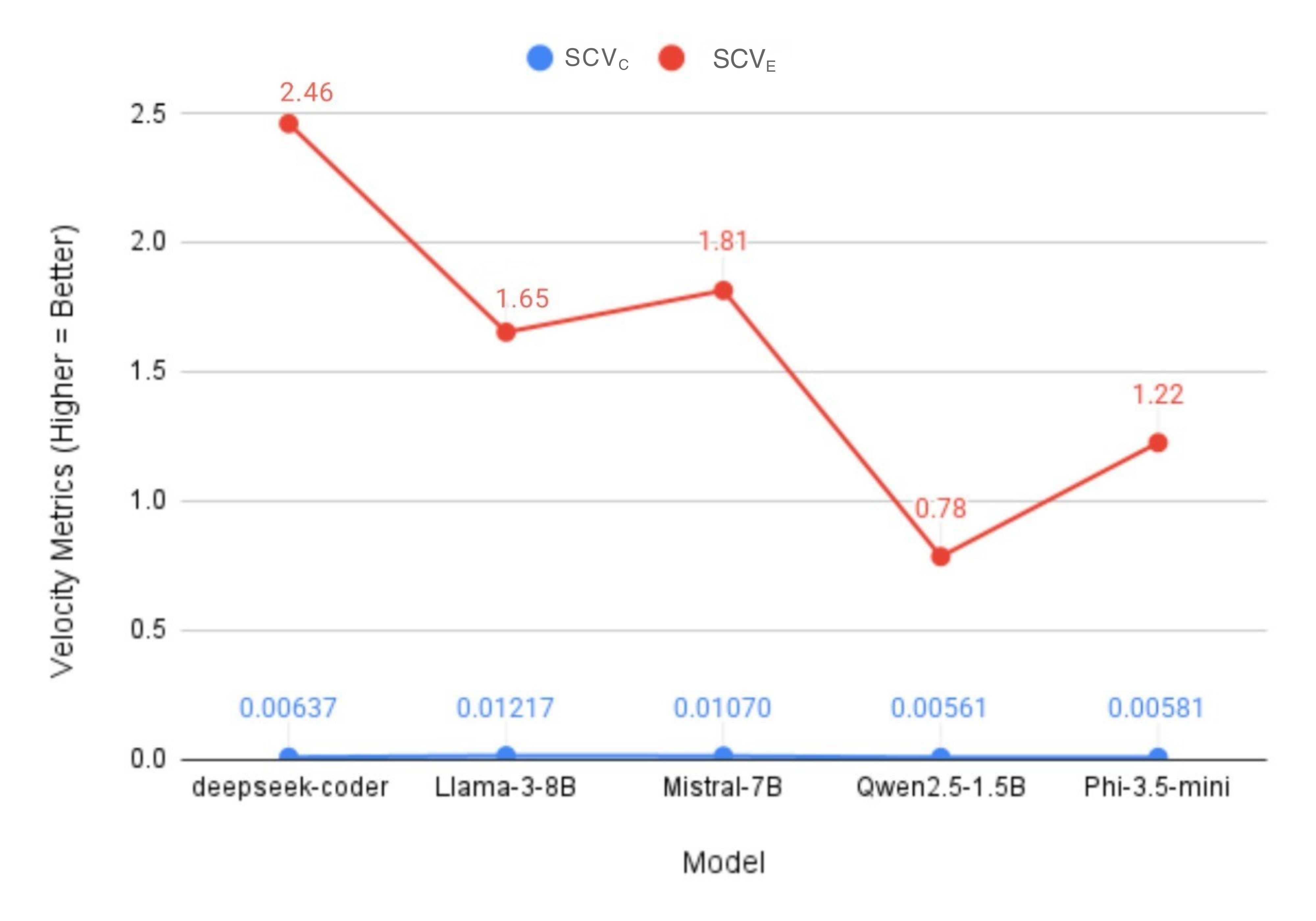}
    \caption{SCV}
    \label{fig:scv}
\end{subfigure}

\hspace{\fill}
\begin{subfigure}{0.32\textwidth}
    \centering
    \includegraphics[width=\linewidth]{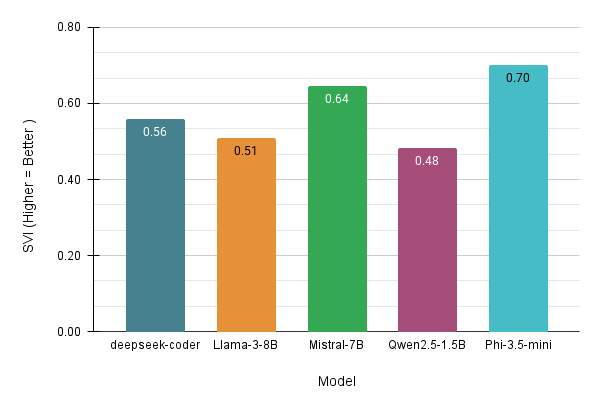}
    \caption{SVI}
    \label{fig:svi}
\end{subfigure}
\hspace{\fill}
\begin{subfigure}{0.32\textwidth}
    \centering
    \includegraphics[width=\linewidth]{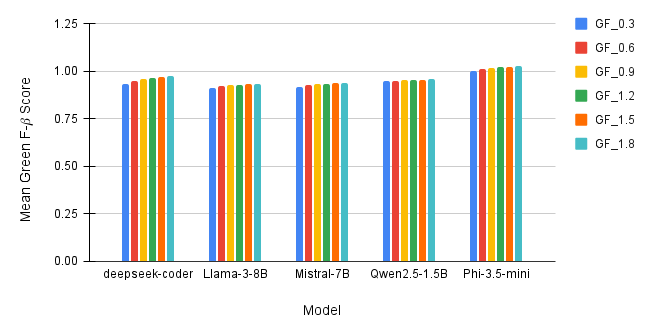}
    \caption{Green F-$\beta$ Score}
    \label{fig:gf_beta}
\end{subfigure}
\hspace{\fill}

\caption{Derived sustainability metrics across SLMs and prompt variants.}
\label{fig:si_gqi_scv_svi_gf}
\end{figure*}

As defined in \eqref{eq:GQI}, \textit{GQI} captures the combined efficiency–quality behavior of each model under adaptive prompting. Figure~\ref{fig:GQI} illustrates GQI values for AP${V_0}$ and AP${V_3}$, where higher scores indicate stronger sustainability–quality alignment.


\textit{Observations:}
\texttt{Qwen2.5-1.5B} achieved the highest overall GQI across both prompt variants (0.0119 at AP${V_0}$ and 0.0101 at AP${V_3}$), demonstrating superior energy–quality balance with minimal degradation under refined prompting. \texttt{Llama-3-8B} and \texttt{Mistral-7B} followed closely, maintaining consistent GQI stability across prompts (0.0108→0.0075 and 0.0099→0.0068, respectively). \texttt{deepseek-coder-7b} showed moderate but reliable performance (0.0088→0.0060), reflecting balanced efficiency–quality behavior. In contrast, \texttt{Phi-3.5-mini} obtained the lowest GQI values (0.0099→0.0088), indicating comparatively weaker sustainability alignment under prompt optimization. These results show that GQI effectively distinguishes models that preserve energy–quality synergy under constrained, emission-aware prompting.

\noindent
\textit{Overall Interpretation:}
The combined \textit{SI} and \textit{GQI} results show that sustainability depends on balancing quality, stability, and energy efficiency rather than minimizing emissions alone. \texttt{Qwen2.5-1.5B} delivers the strongest overall stability and energy–quality alignment, while \texttt{Mistral-7B} excels in emission efficiency and \texttt{Llama-3-8B} maintains steady eco-performance. \texttt{deepseek-coder-7b} provides consistent, low-variance behavior, whereas \texttt{Phi-3.5-mini} highlights the challenges of prompt adaptation, underscoring the need for controlled emissions and reliable output.

\subsubsection{Sustainable Coverage Velocity (SCV$_C$ and SCV$_E$)}
As defined in Eq.(~\ref{eq:scv}), \textit{SCV$_C$} and \textit{SCV$_E$} measure the coverage achieved per unit of time and resource consumption.  
Higher values indicate faster and cleaner execution efficiency. The results are illustrated in Fig.~\ref{fig:scv}

\textit{Observations:}  
\texttt{Llama-3-8B} recorded the highest \textit{SCV$_C$} (0.01217), followed by \texttt{Mistral-7B} (0.01070), indicating efficient time to coverage conversion.  
\texttt{deepseek-coder-7b} achieved moderate coverage velocity (0.00637), showing steady though less rapid computation cycles, while \texttt{Qwen2.5-1.5B} and \texttt{Phi-3.5-mini} remained on the lower side ($\text{SCV$_C$} \leq 0.006$), reflecting comparatively slower throughput under identical prompt conditions.

In contrast, for \textit{SCV$_E$}, which integrates energy consumption, \texttt{deepseek-coder-7b} exhibited the highest value (2.46), marking it as the most energy-efficient and rapid in translating consumed power into coverage.  
\texttt{Mistral-7B} (1.81) and \texttt{Llama-3-8B} (1.65) followed closely, both maintaining strong energy–velocity balance.  
\texttt{Phi-3.5-mini} achieved moderate \textit{SCV$_E$} (1.22), while \texttt{Qwen2.5-1.5B} remained the least energy-productive (0.78), reflecting slower performance relative to power utilization.

\vspace{2pt}
\noindent
\textit{Overall Interpretation:}  
Models with $\text{SCV}_E \geq 1.5$ demonstrate high sustainable velocity, effectively converting both time and energy into functional coverage.  
\texttt{deepseek-coder-7b} leads in overall energy-normalized velocity, while \texttt{Llama-3-8B} and \texttt{Mistral-7B} excel in time-bound efficiency.  
These findings validate that prompt-optimized small language models can achieve not only emission reduction but also higher temporal–energetic throughput—translating directly to faster and greener software test generation.


\subsubsection{Sustainable Velocity Index (SVI)}
As defined in Eq.(~\ref{eq:svi}), \textit{SVI} combines coverage, emission, runtime, and stability into a single normalized measure of time-aware sustainability. The results are illustrated in Fig.~\ref{fig:svi}

\textit{Observations:}  
\texttt{Phi-3.5-mini} achieved the highest \textit{SVI} (0.70), signifying the most balanced performance across time, stability, and carbon efficiency.  
\texttt{Mistral-7B} followed closely (0.64), showing strong sustainability equilibrium between execution time and emission consistency.  
\texttt{deepseek-coder-7b} maintained steady efficiency (0.56), reflecting a reliable though slightly slower balance between energy use and stability.  
\texttt{Llama-3-8B} achieved a moderate score (0.51), while \texttt{Qwen2.5-1.5B} remained the lowest (0.48), indicating higher variability and slower coverage progression.  

\vspace{2pt}
\noindent
\textit{Overall Interpretation:}  
Models with $\text{SVI} \geq 0.6$ exhibit superior temporal–sustainability synergy, balancing carbon intensity, runtime, and consistency effectively.  
\texttt{Phi-3.5-mini} and \texttt{Mistral-7B} stand out as the most sustainably balanced models, while \texttt{deepseek-coder-7b} shows dependable eco-performance with moderate stability.  
These findings affirm that time-aware sustainability depends not only on emission control but also on harmonizing speed and stability to achieve reproducible and energy-conscious model behavior.




\begin{table*}[!ht]
\centering
\scriptsize
\setlength{\tabcolsep}{9pt}
\renewcommand{\arraystretch}{1.1}
\caption{Cross-model qualitative comparison across adaptive prompt versions (AP$_{V_0}$–AP$_{V_3}$) including overall time-aware sustainability index.}
\label{tab:discussion_comparison_final}
\begin{tabular}{|l|p{1cm}|p{1.2cm}|p{1cm}|p{1cm}|p{7.8cm}|}
\hline
\textbf{Metric} & \textbf{AP$_{V_0}$} & \textbf{AP$_{V_1}$} & \textbf{AP$_{V_2}$} & \textbf{AP$_{V_3}$} & \textbf{Interpretation / Dominant Model(s)} \\ \hline
\textbf{SCI (↓)} & Moderate & Slight variance & Lower & Lowest & \texttt{deepseek-coder-7b} achieved minimum emission per run. \\ \hline
\textbf{SEI (↑)} & Moderate & Higher & High & Peak & \texttt{deepseek-coder-7b} showed strongest emission-to-output efficiency; \texttt{Mistral-7B} close second. \\ \hline
\textbf{CER (↑)} & Moderate & Higher & High & Peak & \texttt{Llama-3-8B} maximized coverage-to-carbon balance. \\ \hline
\textbf{SI (↑)} & Stable & Slight improvement & High & Highest & \texttt{Qwen2.5-1.5B} most consistent; \texttt{deepseek-coder-7b} stable across metrics. \\ \hline
\textbf{GQI (↑)} & Moderate & Improving & Higher & Peak & \texttt{Qwen2.5-1.5B} provides the strongest stability and quality alignment, although it lags in time-aware efficiency compared to other models.\\ \hline
\textbf{SCV$_C$ / SCV$_E$ (↑)} & Moderate & Higher & Faster & Fastest & \texttt{Llama-3-8B} led in SCV$_C$; \texttt{deepseek-coder-7b} highest in SCV$_E$. \\ \hline
\textbf{SVI (↑)} & Moderate & Higher & High & Peak & \texttt{Phi-3.5-mini} achieved highest balance of carbon, time, and stability. \\ \hline
\textbf{GF$_\beta$ ($\beta<1$) (↑)} & Moderate & Higher & High & Peak & \texttt{Phi-3.5-mini} and \texttt{Qwen2.5-1.5B} lead in eco-efficiency regime. \\ \hline
\textbf{GF$_\beta$ ($\beta>1$) (↑)} & Moderate & Higher & High & Peak & \texttt{Phi-3.5-mini-instruct} sustained best quality–sustainability trade-off. \\ \hline
\end{tabular}
\end{table*}

\subsubsection{Green F-$\beta$ Score (GF$_\beta$)}
As defined in Eq.(~\ref{eq:gfbeta}), \textit{GF$_\beta$} quantifies the equilibrium between eco-efficiency and output quality, where $\beta<1$ emphasizes sustainability and $\beta>1$ prioritizes quality retention. The results are illustrated in Fig.~\ref{fig:gf_beta}

\textit{Observations:}  
Across both domains, \texttt{Phi-3.5-mini} achieved the highest overall \textit{GF$_\beta$} values, rising from 0.9997 ($\beta=0.3$) to 1.0260 ($\beta=1.8$), indicating superior performance stability and balanced energy–quality trade-offs.  
\texttt{Qwen2.5-1.5B} followed with consistently high scores (0.9465 → 0.9571), showing excellent alignment between energy optimization and output accuracy.  
\texttt{deepseek-coder-7b} recorded a steady eco–quality balance (0.9351 → 0.9728), reflecting strong adaptability across both regimes.  
\texttt{Mistral-7B} maintained stable mid-range values (0.9189 → 0.9378), while \texttt{Llama-3-8B} displayed slightly lower but consistent progression (0.9130 → 0.9346).  

The comparative plot (Fig.~\ref{fig:gf_beta}) illustrates that all models maintain near-uniform improvement as $\beta$ increases, with \texttt{Phi-3.5-mini} showing the steepest upward trajectory, signifying stronger quality retention without compromising eco-efficiency.


\noindent
\textit{Overall Interpretation:}  
Models with $\text{GF}_\beta \geq 0.95$ exhibit excellent harmony between sustainability and performance.  
\texttt{Phi-3.5-mini} demonstrates the strongest balance between energy optimization and accuracy, while \texttt{Qwen2.5-1.5B} and \texttt{deepseek-coder-7b} show highly consistent energy–quality coupling across all $\beta$ domains.  
Together, these results reinforce that sustainable AI behavior extends beyond emission reduction—requiring models to sustain proportional quality gains even under constrained energy budgets, ensuring dependable and environmentally aligned inference performance.
\\
\noindent

The comparative trends in Table~\ref{tab:discussion_comparison_final} show that model sustainability dynamics evolve distinctly across progressive prompt versions.
\texttt{Phi-3.5-mini} demonstrates the highest overall sustainability maturity, combining balanced emission control, execution stability, and eco–quality efficiency.
\texttt{Mistral-7B} follows closely, exhibiting strong energy–quality synergy and consistent time-aware performance.
\texttt{deepseek-coder-7b} remains the most carbon- and energy-efficient, maintaining reliable performance across velocity-based metrics.
\texttt{Llama-3-8B} sustains balanced velocity and coverage optimization, while \texttt{Qwen2.5-1.5B} provides the strongest stability and quality alignment, although it lags in time-aware efficiency.
Overall, no single model dominates across all sustainability dimensions—the optimal selection depends on target priorities such as emission control, stability, or eco–quality trade-offs within sustainability-constrained inference environments.

\section{Threats to Validity}
\label{sec:threats}

This study presents several potential threats to validity.

\textit{Internal Validity:} Although the experimental pipeline was kept constant, uncontrolled factors such as system background processes, variability in model loading, or cache effects may have influenced the energy measurements and execution time, potentially introducing bias in reported outcomes.

\textit{External Validity:} The generalizability of the results is limited due to the reliance on a single benchmark (HumanEval), fixed hardware configuration, and specific assumptions of the carbon intensity of electricity. Performance and emission outcomes may vary across other datasets, programming tasks, geographic regions (owing to electricity grid differences), or hardware architectures such as TPU, AMD GPU, or edge devices.

\textit{Construct Validity:} Emission values were derived using estimated carbon intensity from CodeCarbon rather than real-time smart-meter measurements. Therefore, the reported results serve as approximations rather than absolute emission values. Similarly, test coverage and accuracy may not fully capture the broader notions of quality, usability, or robustness of AI-generated test scripts.

\textit{Statistical Conclusion Validity:} Experiments were carried out with a limited number of trials per model–prompt pair, which may affect the reliability of observed trends in emission variations and time–efficiency relationships. Minor fluctuations might not reach statistical significance, and potential correlations should be interpreted with caution.

\section{Conclusion}\label{sec:Conclusion}
This study presented a comprehensive sustainability evaluation framework for small language models (SLMs) using the HumanEval dataset and progressive prompt variants (AP$_{V_0}$–AP$_{V_3}$). By integrating primary and derived sustainability metrics—SCI, SEI, CER, SI, GQI, SCV, SVI, and GF$_\beta$—the framework quantified not only the energy and emission footprints of model inference but also their stability, temporal efficiency, and quality trade-offs.

Experimental analysis revealed model-specific strengths: \texttt{deepseek-coder-7b} excelled in emission and energy efficiency, \texttt{Phi-3.5-mini} provided the best eco–quality and temporal balance, and \texttt{Qwen2.5-1.5B} showed the highest run-to-run stability. These results highlight the multidimensional nature of sustainability—no single model performs best across all axes, and the optimal choice depends on desired priorities such as emission control, stability, or output quality.

Future work will refine prompt design for greener configurations, extend the framework to additional model families, and evaluate unquantized executions and diverse hardware settings to uncover deeper sustainability patterns.

\bibliographystyle{IEEEtran}
\bibliography{IEEEfull}

\begin{thebibliography}{10}
\providecommand{\url}[1]{#1}
\csname url@samestyle\endcsname
\providecommand{\newblock}{\relax}
\providecommand{\bibinfo}[2]{#2}
\providecommand{\BIBentrySTDinterwordspacing}{\spaceskip=0pt\relax}
\providecommand{\BIBentryALTinterwordstretchfactor}{4}
\providecommand{\BIBentryALTinterwordspacing}{\spaceskip=\fontdimen2\font plus
\BIBentryALTinterwordstretchfactor\fontdimen3\font minus \fontdimen4\font\relax}
\providecommand{\BIBforeignlanguage}[2]{{%
\expandafter\ifx\csname l@#1\endcsname\relax
\typeout{** WARNING: IEEEtran.bst: No hyphenation pattern has been}%
\typeout{** loaded for the language `#1'. Using the pattern for}%
\typeout{** the default language instead.}%
\else
\language=\csname l@#1\endcsname
\fi
#2}}
\providecommand{\BIBdecl}{\relax}
\BIBdecl

\bibitem{zadenoori2025model}
M.~A. Zadenoori, V.~D. Martino, J.~Dąbrowski, X.~Franch, and A.~Ferrari, ``Does model size matter? a comparison of small and large language models for requirements classification,'' \emph{arXiv preprint arXiv:2510.21443}, 2025.

\bibitem{belcak2025slms}
P.~Belcak, G.~Heinrich, S.~Diao, Y.~Fu, X.~Dong, S.~Muralidharan, Y.~C. Lin, and P.~Molchanov, ``Small language models are the future of agentic ai,'' \emph{arXiv preprint arXiv:2506.02153}, 2025.

\bibitem{ashraf2025energyaware}
H.~Ashraf, S.~M. Danish, A.~Leivadeas, Y.~Otoum, and Z.~Sattar, ``Energy-aware code generation with llms: Benchmarking small vs. large language models for sustainable ai programming,'' in \emph{Proceedings of the 47th International Conference on Software Engineering (ICSE)}.\hskip 1em plus 0.5em minus 0.4em\relax IEEE, 2025.

\bibitem{zhang2025llmunittesting}
Q.~Zhang, C.~Fang, S.~Gu, Y.~Shang, Z.~Chen, and L.~Xiao, ``Large language models for unit testing: A systematic literature review,'' \emph{arXiv preprint arXiv:2506.15227}, 2025.

\bibitem{ma2024prompt}
W.~Ma, C.~Yang, and C.~Kästner, ``(why) is my prompt getting worse? rethinking regression testing for evolving llm apis,'' in \emph{Proceedings of the Conference on AI Engineering: Software Engineering for AI (CAIN)}.\hskip 1em plus 0.5em minus 0.4em\relax ACM, 2024, pp. 1--6.

\bibitem{almonte2025automated}
J.~T. Almonte, S.~A. Boominathan, and N.~Nascimento, ``Automated non-functional requirements generation in software engineering with large language models: A comparative study,'' \emph{arXiv preprint arXiv:2503.15248}, 2025.

\bibitem{wu2023greenprompting}
E.~Z. Wu, J.~Gonzalez, Y.~Goldberg, and T.~Hashimoto, ``Green prompting: Reducing inference-time emissions of language models,'' \emph{arXiv preprint arXiv:2305.17138}, 2023.

\bibitem{cappendijk2025generating}
T.~Cappendijk, P.~de~Reus, and A.~Oprescu, ``An exploration of prompting llms to generate energy-efficient code,'' in \emph{Proceedings of the 47th International Conference on Software Engineering: GREENs Track}.\hskip 1em plus 0.5em minus 0.4em\relax Ottawa, Canada: ACM, 2025.

\bibitem{oprescu2023prompt}
A.~Oprescu, P.~Reus, and T.~Cappendijk, ``Prompt engineering for energy efficiency: Lessons from code generation,'' in \emph{Proceedings of the 2023 International Conference on Green Software}.\hskip 1em plus 0.5em minus 0.4em\relax ACM, 2023.

\bibitem{deng2024llmcarbon}
C.~Deng, Y.~Liu, R.~Wang, Y.~Wen, Y.~Meng, S.~Li, H.~Wang, X.~Jin, and Y.~Xu, ``Llmcarbon: Quantifying and mitigating the carbon footprint of large language models,'' \emph{arXiv preprint arXiv:2402.08776}, 2024.

\bibitem{luccioni2023ml}
A.~S. Luccioni, E.~Cornblath, L.~Kaack, A.~Kapur, and A.~Lacoste, ``Ml co2 emissions tracker: Estimating the carbon footprint of ml models in real time,'' \emph{arXiv preprint arXiv:2302.08476}, 2023.

\bibitem{li2023energy}
Y.~Li, R.~Chen, M.~Gupta, and D.~Rao, ``Energy consumption patterns in automated test script generation,'' in \emph{Proceedings of the 45th International Conference on Software Engineering (ICSE)}.\hskip 1em plus 0.5em minus 0.4em\relax IEEE, 2023, pp. 845--856.

\bibitem{sharma2023sustainable}
R.~Sharma, N.~Batra, and X.~Liu, ``Sustainable test automation: Quantifying the environmental impact of code testing with slms,'' \emph{Journal of Software Sustainability}, vol.~15, no.~3, pp. 225--239, 2023.

\bibitem{fraser2011evosuite}
G.~Fraser and A.~Arcuri, ``Evosuite: Automatic test suite generation for object-oriented software,'' in \emph{Proceedings of the 19th ACM SIGSOFT Symposium and the 13th European Conference on Foundations of Software Engineering}.\hskip 1em plus 0.5em minus 0.4em\relax ACM, 2011, pp. 416--419.

\bibitem{lago2014systematic}
\BIBentryALTinterwordspacing
P.~Lago, Q.~Gu, and P.~Bozzelli, ``A systematic literature review of green software metrics,'' \emph{VU University Amsterdam Technical Report}, 2014. [Online]. Available: \url{https://research.vu.nl/en/publications/a-systematic-literature-review-of-green-software-metrics}
\BIBentrySTDinterwordspacing

\bibitem{mourao2018green}
B.~C. Mourão, L.~Karita, and I.~do~Carmo~Machado, ``Green and sustainable software engineering: A systematic mapping study,'' in \emph{Proceedings of the SBQS}.\hskip 1em plus 0.5em minus 0.4em\relax ACM, 2018, pp. 1--10.

\bibitem{kapoor2024green}
S.~Kapoor, ``Green software quality: A comprehensive framework for sustainable metrics in software development,'' \emph{International Journal of Computer Trends and Technology}, vol.~72, no.~10, pp. 113--120, 2024.

\bibitem{iyer2023cloud}
V.~H. Iyer, R.~Pathak, A.~Sridhar, V.~Apte, R.~Gopalakrishnan, D.~Kalathil, and P.~Ranganathan, ``Carbon-aware large language model inference in the cloud,'' \emph{arXiv preprint arXiv:2309.08101}, 2023.

\bibitem{verdecchia2021green}
R.~Verdecchia, P.~Lago, C.~Ebert, and C.~de~Vries, ``Green it and green software,'' \emph{IEEE Software}, vol.~38, no.~6, pp. 7--15, 2021.

\bibitem{vartziotis2024learn}
T.~Vartziotis, I.~Dellatolas, G.~Dasoulas, M.~Schmidt, F.~Schneider, T.~Hoffmann, S.~Kotsopoulos, and M.~Keckeisen, ``Learn to code sustainably: An empirical study on green code generation,'' in \emph{Proceedings of the 2024 International Workshop on Large Language Models for Code (LLM4Code)}.\hskip 1em plus 0.5em minus 0.4em\relax ACM, 2024, pp. 1--8.

\bibitem{gsf2025}
{Green Software Foundation}, ``Green software foundation website,'' \emph{Green Software Foundation}, 2021, available online: https://greensoftware.foundation.

\bibitem{strubell2019energy}
E.~Strubell, A.~Ganesh, and A.~McCallum, ``Energy and policy considerations for deep learning in nlp,'' in \emph{Proceedings of the 57th Annual Meeting of the Association for Computational Linguistics (ACL)}.\hskip 1em plus 0.5em minus 0.4em\relax Association for Computational Linguistics, 2019, pp. 3645--3650.

\bibitem{rashkin2021words}
H.~Rashkin, M.~Sap, M.~Forbes, and Y.~Choi, ``Words to watts: A benchmark for measuring the energy efficiency of nlp models and tasks,'' in \emph{Proceedings of the 2021 Conference on Empirical Methods in Natural Language Processing (EMNLP)}, 2021, pp. 10\,000--10\,015.

\end{thebibliography}

\end{document}